\documentclass[aps,preprint]{revtex4}%
\usepackage{amsfonts}
\usepackage{amsmath}
\usepackage{amssymb}
\usepackage{graphicx}%
\providecommand{\U}[1]{\protect\rule{.1in}{.1in}}
\newtheorem{theorem}{Theorem}

\newtheorem{remark}[theorem]{Remark}

\begin{document}
\preprint{ }
\title[Short title for running header]{Dynamical breaking of symmetries beyond the standard model and supergeometry }
\author{Andrej B. Arbuzov}
\affiliation{Joint Institute for Nuclear Research, Dubna, 141980, Russia}
\affiliation{Dubna State University, Dubna, 141980, Russia}
\author{Diego Julio Cirilo-Lombardo}
\affiliation{CONICET-Universidad de Buenos Aires, Instituto de Fisica del Plasma (INFIP),
Buenos Aires, Argentina}

\begin{abstract}
Group theoretical realizations containing the electroweak sector of the
Standard Model are discussed from the supersymmetry point of view.
Dynamical breaking of the symmetry is
performed and the corresponding quadratic (super Yang-Mills) Lagrangian is
obtained. Supercoherent states of the Klauder-Perelomov type are defined to
enlarge the symmetry taking into account the geometry of the coset based in
the simplest supergroup $SU\left(2\mid1\right)$ as the structural basis of 
the electroweak sector of the SM. The extended model is superintegrable and 
the superconnection in the odd part takes a dynamical character. The physical 
and geometrical implications of the additional degrees of freedom interpreted
as a hidden sector of the representation are briefly discussed.
\end{abstract}

\eid{identifier}





\maketitle
\tableofcontents

\section{Introduction}

As it is well known, the Standard Model (SM) of particle physics and General
Relativity are extremely successful phenomenologically based theories. In the
case of the SM, the apparent last building block associated with the Higgs
sector has been detected experimentally in 2012. However, the discovery of the
Higgs boson with the mass of about $125$~GeV doesn't solve the hierarchy
problem of the SM. There are also several questions coming from both
astrophysical and cosmological scenarios where the necessity of new exotic
states (possibly from supersymmetrical or composite) is quite obvious. This
incompleteness of the SM, which is the case to be particularly treated here,
is a clear indication that new models with a richer geometric and algebraic
structure should be considered. With this aim and motivation in this paper we
will apply the model described in our previous works to solve the problem of
the symmetry rupture as the content of fields, both those corresponding to the
SM, plus other additional ones emerging from the group structure.

In the language of superbundles in our previous work~\cite{Arbuzov:2017frk},
the fiber was defined as $E=E(U_{4}$, $F=E^{2}(V_{4}^{\prime},S)$,
$G=SO(4,2))$ where $P=P(U_{4}$, $F=G=SO(4,2))$ was the associated bundle and
$G$ can be viewed as the bundle $P(G/H,H)$. The geometrical meaning in this
statement, is that the soldering of $E$ is obtained by identifying the tangent
space of $V_{4}^{\prime}$ of $F$ at the point, namely, $\psi=\psi^{0} $ with
the local tangent space $T_{x}$ of $U_{4}$ at $x$ through an isomorphism.
Consequently, the point $\psi^{0}$ $\left(  \text{the origin of }V_{4}%
^{\prime}\right)  $ is geometrically the contact point between the base space
(spacetime) and the fiber at $x\in U_{4}$.

Now we treat the simplest supergroup as the principal bundle $SU\left(
2|1\right)  $ that have several very interesting features. In particular, the
matrix representation is performed exactly as the $SU\left(  3\right)  $
Gell-Mann matrices with the only change in the structure of the diagonal
matrix $\Lambda_{8}$ that is of non-zero trace. This important characteristic
of the supergroup $SU\left(  2|1\right)  $ facilitates calculations (which are
usually rather complex, in particular for the construction of supercoherent
states). Moreover, the interpretation of the results can be performed from the
point of view of a grand unification theory beyond the
SM~\cite{Cirilo-Lombardo:2015fua}.

\section{Gauge structure and superconnections: $SU(2/1)$}

The superconnection was introduced by Quillen in mathematics, it is a
supermatrix belonging to a given supergroup $S$ valued over elements of a
Grassmann algebra of forms. The even part of the supermatrix is valued over
the gauge potentials of the even subgroup $G\subset S$, one-forms $Bdx$ on the
base $M$ manifold of the bundle, realizing the \textquotedblleft
gauging\textquotedblright\ of $G$. The odd part of the supermatrix,
representing the quotient $S/G=H\subset S$, is valued over zero-forms in that
Grassmann algebra, physically interpreted as the Higgs multiplet in a
spontaneously broken $G$ gauge theory. In some quantum treatments which are
set to reproduce geometrically the ghost fields and BRST equations, the
Grassmann algebra is taken over the complete bundle variable.

The first physical example of a superconnection preceded Quillen's theory.
This was the Neeman's et al. supergroup
proposal~\cite{Neeman:2005dbs,Fairlie:1979at} for an algebraically irreducible
electroweak unification. Lacking Quillen's generalized formulation, the model
appeared to suffer from spin-statistics interpretative complications for
physical fields. The structural $Z$ grading of Lie superalgebras, as
previously used in physics (e.g. in SUGRA etc.) corresponds to the grading
inherent in quantum statistics, i.e. to Bose--Fermi transitions, so that the
invariance under the supergroup represents symmetry between bosons and
fermions. In the proposal~\cite{Neeman:2005dbs}, however, though the
superconnection itself does fit the quantum statistics ansatz. This is
realized through the order of the forms in the geometrical space of the
Grassmann algebra, rather than through the quantum statistics of the particle
Hilbert space.

\subsection{Supergroup structure}

The generator structure for exponential representation of the group under
considerations reads~\cite{Neeman:2005dbs,Bars:1982ep,Arbuzov:2017frk}
\begin{equation}
J=\left(
\begin{array}
[c]{cc}%
M & \phi\\
\overline{\phi} & N
\end{array}
\right)  .
\end{equation}

The even generators in this structure are
\begin{equation}
\lambda_{1}=\left(
\begin{array}
[c]{ccc}%
0 & 1 & 0\\
1 & 0 & 0\\
0 & 0 & 0
\end{array}
\right)  ,\quad\lambda_{2}=\left(
\begin{array}
[c]{ccc}%
0 & -i & 0\\
i & 0 & 0\\
0 & 0 & 0
\end{array}
\right)  ,\quad\lambda_{3}=\left(
\begin{array}
[c]{ccc}%
1 & 0 & 0\\
0 & -1 & 0\\
0 & 0 & 0
\end{array}
\right)  ,\quad\lambda_{8}=\left(
\begin{array}
[c]{ccc}%
-1 & 0 & 0\\
0 & -1 & 0\\
0 & 0 & -2
\end{array}
\right)
\end{equation}
and the odd ones are
\begin{equation}
\lambda_{4}=\left(
\begin{array}
[c]{ccc}%
0 & 0 & 1\\
0 & 0 & 0\\
1 & 0 & 0
\end{array}
\right)  ,\quad\lambda_{5}=\left(
\begin{array}
[c]{ccc}%
0 & 0 & -i\\
0 & 0 & 0\\
i & 0 & 0
\end{array}
\right)  ,\quad\lambda_{6}=\left(
\begin{array}
[c]{ccc}%
0 & 0 & 0\\
0 & 0 & 1\\
0 & 1 & 0
\end{array}
\right)  ,\quad\lambda_{7}=\left(
\begin{array}
[c]{ccc}%
0 & 0 & 0\\
0 & 0 & -i\\
0 & i & 0
\end{array}
\right)  .
\end{equation}
The diagonal matrices are related to the following important operators:
$I_{L}^{3}\equiv\frac{1}{2}\lambda_{3}$, the gradation operator $U_{Q}%
\equiv\sqrt{3}\lambda_{8}$, and the new operator that is fixed by the
orthogonality in $U(2/1)$:
\begin{equation}
\lambda_{0}\equiv\sqrt{\frac{2}{3}}\left(
\begin{array}
[c]{ccc}%
1 & 0 & 0\\
0 & 1 & 0\\
0 & 0 & -1
\end{array}
\right)  .
\end{equation}
In previous studies the latter was related with the chirality of the model
defined by $\sqrt{\frac{3}{2}}\lambda_{0}$.

\subsection{Element of the supergroup: $\mathcal{K\in B}_{1},\mathcal{H\in}$
$\mathcal{B}_{0}:$}

We have seen that in the ordinary parameterization, an element of the
supergroup can be written
as~\cite{Bars:1982ep,Cirilo-Lombardo:2015fua,Cirilo-Lombardo:2016ogj}
\begin{equation}
\mathcal{K}=\frac{1}{\sqrt{1+\phi^{\dagger}\phi}}\left(
\begin{array}
[c]{cc}%
\mathbb{I}_{2\times2} & \phi\\
\phi^{\dagger} & 1
\end{array}
\right)
\end{equation}
with
\begin{equation}
\phi=\frac{\tan\sqrt{v^{\dagger}v}}{\sqrt{v^{\dagger}v}}v\text{ \ \ \ and
\ \ \ \ }v=\left(
\begin{array}
[c]{c}%
\theta_{4}-i\theta_{5}\\
\theta_{6}-i\theta_{7}%
\end{array}
\right)
\end{equation}
and
\begin{equation}
\mathcal{H}=\left(
\begin{array}
[c]{cc}%
we^{-i\theta_{8}/\sqrt{3}} & \mathbf{0}_{2\times1}\\
\mathbf{0}_{1\times2} & e^{-2i\theta_{8}/\sqrt{3}}%
\end{array}
\right)
\end{equation}%
\begin{equation}
\mathcal{U}=\mathcal{KH},
\end{equation}
with
\begin{align}
w  &  =\frac{i}{\sqrt{1+W^{2}}}\left\{  \overset{-i\mathbb{I}}{\overbrace
{\left(
\begin{array}
[c]{cc}%
-i & 0\\
0 & -i
\end{array}
\right)  }}+\overset{Y\cdot\sigma}{\overbrace{\left(
\begin{array}
[c]{cc}%
W_{3} & W_{-}\\
W_{+} & -W_{3}%
\end{array}
\right)  }}\right\}  ,\text{ \ \ \ \ \ \ \ \ }W_{3,\pm}=\frac{\tan\sqrt
{\theta^{2}}}{\sqrt{\theta^{2}}}\theta_{3,\pm},\text{ \ \ \ \ \ }\\
&  \text{\ }%
\begin{array}
[c]{c}%
\text{\ }\theta^{2}=\theta_{1}^{2}+\theta_{2}^{2}+\theta_{3}^{2},\qquad
\theta_{-}=\left(  \theta_{1}-i\theta_{2}\right)  ,\qquad\theta_{+}=\left(
\theta_{1}+i\theta_{2}\right)  .
\end{array}
\end{align}%
\begin{equation}
\mathcal{U}=\frac{1}{\sqrt{1+Z^{\dagger}Z}}\left(
\begin{array}
[c]{cc}%
we^{-i\theta_{8}/\sqrt{3}} & \phi e^{-2i\theta_{8}/\sqrt{3}}\\
\phi^{\dagger}we^{-i\theta_{8}/\sqrt{3}} & e^{-2i\theta_{8}/\sqrt{3}}%
\end{array}
\right)  .
\end{equation}
Consequently, the element of the supergroup can be explicitly written as
\begin{equation}
\mathcal{U}=\frac{1}{\sqrt{1+Z^{\dagger}Z}}\left(
\begin{array}
[c]{cc}%
we^{-i\theta_{8}/\sqrt{3}} & \phi e^{-2i\theta_{8}/\sqrt{3}}\\
\phi^{\dagger}we^{-i\theta_{8}/\sqrt{3}} & e^{-2i\theta_{8}/\sqrt{3}}%
\end{array}
\right)  .
\end{equation}

Note that \newline i) in $\mathcal{U}$, the even part of the group structure
(corresponding with the electroweak sector, namely $SU(2)\otimes U(1))$ is
exactly preserved; \newline ii) the non diagonal blocks (the odd part) in
$\mathcal{U}$ can be interpreted as new fermion-boson interaction; \newline
iii) $\mathcal{H=}\exp\left(  \underset{k=1,2,3,8}{\sum\theta_{k}\sigma_{k}%
}\right)  $ and $w\mathcal{=}\exp\left(  \underset{k=1,2,3}{\sum\theta
_{k}\sigma_{k}}\right)  \in SU(2)_{L}$; \newline iv) there are other
parameterization \textit{symmetrical\ (Borel type)} ones involving ladder
operators that we will use in the construction of the corresponding
supercoherent states.

\section{Supercurvature and Lagrangian}

Due to the identification of the superconnection and $\mathcal{U}$, we can see
that from
\begin{align}
\Gamma &  =\frac{e^{-i\theta_{8}/\sqrt{3}}}{\sqrt{1+\phi^{\dagger}\phi}%
}\left\{  \left[
\begin{array}
[c]{cc}%
w & \mathbf{0}_{2\times1}\\
\mathbf{0}_{1\times2} & e^{-i\theta_{8}/\sqrt{3}}%
\end{array}
\right]  +\left[
\begin{array}
[c]{cc}%
\mathbf{0}_{2\times2} & \phi e^{-i\theta_{8}/\sqrt{3}}\\
\phi^{\dagger}w & 0
\end{array}
\right]  \right\} \\
&  \equiv\Gamma_{even}+\Gamma_{odd}%
\end{align}
one can obtain the super-Riemannian curvature in the language of superforms:
\begin{align}
d\Gamma_{even}+\Gamma_{even}\wedge\Gamma_{even}+\Gamma_{odd}\vee\Gamma_{odd}
&  \rightarrow d\Gamma_{even}+\left[  \Gamma_{even},\Gamma_{even}\right]
+\left\{  \Gamma_{odd},\Gamma_{odd}\right\} \\
d\Gamma_{odd}+\Gamma_{even}\wedge\Gamma_{odd}  &  \rightarrow d\Gamma
_{even}+\left[  \Gamma_{even},\Gamma_{odd}\right]  .
\end{align}
To compute the above equations that define the supercurvature, it is useful to
have in mind that the tensor product of a commutative superalgebra of
differential forms and a Lie superalgebra is again a Lie superalgebra with the
product
\begin{equation}
\left[  a\otimes X,b\otimes Y\right]  =-1^{\left\vert X\right\vert \left\vert
b\right\vert }\left(  a\wedge b\right)  \otimes\left[  X,Y\right]  .
\end{equation}
The next step is to compute the above quantities explicitly. To this end we
rewrite the connection in the original form as
\begin{gather}
\Gamma=\underset{\Omega}{\underbrace{\frac{e^{-i\theta_{8}/\sqrt{3}}}%
{\sqrt{1+\phi^{\dagger}\phi}}}}\left\{  \underset{\widetilde{\Gamma}_{even}%
}{\underbrace{\left[
\begin{array}
[c]{cc}%
\overset{w}{\overbrace{\frac{i}{\sqrt{1+W^{2}}}\left[
\begin{array}
[c]{cc}%
W_{3}-i & W_{-}\\
W_{+} & -W_{3}-i
\end{array}
\right]  }} & \mathbf{0}_{2\times1}\\
\mathbf{0}_{1\times2} & e^{-i\theta_{8}/\sqrt{3}}%
\end{array}
\right]  }}+\underset{\widetilde{\Gamma}_{odd}}{\underbrace{i\left[
\begin{array}
[c]{cc}%
\mathbf{0}_{2\times2} & \phi\\
\phi^{\dagger} & 0
\end{array}
\right]  \widetilde{\Gamma}_{even}},}\right\} \\
=\frac{e^{-i\theta_{8}/\sqrt{3}}}{\sqrt{1+Z^{\dagger}Z}}\left\{  \left(
1+i\left[
\begin{array}
[c]{cc}%
\mathbf{0}_{2\times2} & \phi\\
\phi^{\dagger} & 0
\end{array}
\right]  \right)  \underset{\widetilde{\Gamma}_{even}}{\underbrace{\left[
\begin{array}
[c]{cc}%
w & \mathbf{0}_{2\times1}\\
\mathbf{0}_{1\times2} & e^{-i\theta_{8}/\sqrt{3}}%
\end{array}
\right]  }}\right\}  ,\\
d\Gamma=d\left(  \frac{e^{-i\theta_{8}/\sqrt{3}}}{\sqrt{1+\phi^{\dagger}\phi}%
}\right)  \wedge\left(  \widetilde{\Gamma}_{even}+\widetilde{\Gamma}%
_{odd}\right)  +\frac{e^{-i\theta_{8}/\sqrt{3}}}{\sqrt{1+\phi^{\dagger}\phi}%
}\left(  d\widetilde{\Gamma}_{even}+d\widetilde{\Gamma}_{odd}\right)  ,\\
dw=d\left(  \frac{i}{\sqrt{1+W^{2}}}\right)  \wedge\left(
\begin{array}
[c]{cc}%
W_{3}-i & W_{-}\\
W_{+} & -W_{3}-i
\end{array}
\right)  +\frac{i}{\sqrt{1+W^{2}}}d\left(
\begin{array}
[c]{cc}%
W_{3}-i & W_{-}\\
W_{+} & -W_{3}-i
\end{array}
\right)  .
\end{gather}
Consequently,
\begin{gather}
d\widetilde{\Gamma}_{even}=\left(
\begin{array}
[c]{cc}%
dw & \mathbf{0}_{2\times1}\\
\mathbf{0}_{1\times2} & d\left(  e^{-i\theta_{8}/\sqrt{3}}\right)
\end{array}
\right)  ,\\
d\widetilde{\Gamma}_{odd}=i\left(
\begin{array}
[c]{cc}%
\mathbf{0}_{2\times2} & d\phi-\phi\wedge d\left(  \ln e^{-i\theta_{8}/\sqrt
{3}}\right) \\
d\phi^{\dagger}+\phi^{\dagger}\wedge d\ln w & 0
\end{array}
\ \ \right)  \widetilde{\Gamma}_{even},
\end{gather}%
\begin{align}
\left[  \widetilde{\Gamma}_{even},\widetilde{\Gamma}_{even}\right]   &
=\frac{-1}{1+W^{2}}\left(
\begin{array}
[c]{cc}%
W_{i}\wedge W_{j}\epsilon_{ijk}\sigma_{k} & \mathbf{0}_{2\times1}\\
\mathbf{0}_{1\times2} & 0
\end{array}
\right)  ,\\
\left\{  \widetilde{\Gamma}_{odd},\widetilde{\Gamma}_{odd}\right\}   &
=-4i\left(
\begin{array}
[c]{cc}%
\phi e^{-i\theta_{8}/\sqrt{3}}\phi^{\dagger} & \mathbf{0}_{2\times1}\\
\mathbf{0}_{1\times2} & \phi^{\dagger}w\phi
\end{array}
\right)  \widetilde{\Gamma}_{even},\\
\left[  \widetilde{\Gamma}_{even},\widetilde{\Gamma}_{odd}\right]   &
=-2i\left(
\begin{array}
[c]{cc}%
\mathbf{0}_{2\times2} & \left(  w-\mathbb{I}e^{-i\theta_{8}/\sqrt{3}}\right)
\phi\\
\phi^{\dagger}\left(  \mathbb{I}e^{-i\theta_{8}/\sqrt{3}}-w\right)  & 0
\end{array}
\right)  \widetilde{\Gamma}_{even}.
\end{align}
The supercurvatures with the above definitions can be computed in the usual way.

\subsection{Odd supercurvature and Weinberg angle}

The explicit computation of the odd supercurvature gives
\begin{align}
&  d\Gamma_{odd}+\Gamma_{even}\wedge\Gamma_{odd}\equiv\nonumber\\
&  \qquad-i\Omega\Gamma_{even}\left\{  \left(
\begin{array}
[c]{cc}%
\mathbf{0}_{2\times2} & d\phi-2\Omega\left(  w-\mathbb{I}e^{-i\theta_{8}%
/\sqrt{3}}\right)  \phi\\
d\phi^{\dagger}+2\Omega\left(  w-\mathbb{I}e^{-i\theta_{8}/\sqrt{3}}\right)
\phi^{\dagger}0 &
\end{array}
\right)  \right. \nonumber\\
&  \qquad+\left.  i\left(
\begin{array}
[c]{cc}%
\mathbf{0}_{2\times2} & -\phi\wedge d\ln\left(  \Omega e^{-i\theta_{8}%
/\sqrt{3}}\right) \\
\phi^{\dagger}\wedge d\ln\left(  \Omega w\right)  & 0
\end{array}
\right)  \right\}  .
\end{align}
Now we use the detailed association (at the end in the Appendix III), to
identify the physical fields, namely
\begin{align}
&  \mathcal{W\rightarrow}\widetilde{\mathcal{W}}\mathcal{\equiv}\widetilde
{W}\cdot\sigma\rightarrow\frac{e^{-i\theta_{8}/\sqrt{3}}}{\sqrt{1+\phi
^{\dagger}\phi}}\frac{i}{\sqrt{1+W^{2}}}\underset{W\cdot\sigma}{\underbrace
{\left(
\begin{array}
[c]{cc}%
W_{3} & W_{-}\\
W_{+} & -W_{3}%
\end{array}
\right)  }},\\
&  -\frac{1}{\sqrt{3}}B\cdot\mathbb{I}_{2\times2}\rightarrow-\frac{1}{\sqrt
{3}}\widetilde{B}\cdot\mathbb{I}_{2\times2}\underset{B-d\theta}{\underbrace
{\rightarrow}}-\frac{e^{-i\theta_{8}/\sqrt{3}}}{\sqrt{1+\phi^{\dagger}\phi}%
}\frac{i}{\sqrt{1+W^{2}}}\mathbb{I}_{2\times2}.
\end{align}
For consistency, it should be supplemented with the transformation law of the
gauge of the 1-form associated with $\lambda_{8}$:
\begin{equation}
B\rightarrow B-d\theta.\nonumber
\end{equation}
Then, rewriting accordingly we arrive to%
\begin{align}
&  d\Gamma_{odd}+\Gamma_{even}\wedge\Gamma_{odd}\nonumber\\
&  \equiv-i\left(
\begin{array}
[c]{cc}%
\mathbf{0}_{2\times2} & d\widetilde{\phi}-2i\left[  \widetilde{W}\cdot
\sigma-\frac{g^{\prime}}{\sqrt{3}}\widetilde{B}\cdot\mathbb{I}_{2\times
2}\right]  \widetilde{\phi}\\
d\widetilde{\phi}^{\dagger}+2i\widetilde{\phi}^{\dagger}\left[  \widetilde
{W}\cdot\sigma-\frac{g^{\prime}}{\sqrt{3}}\widetilde{B}\cdot\mathbb{I}%
_{2\times2}\right]  & 0
\end{array}
\right)  ,
\end{align}
where we defined $g^{\prime}\equiv\left(  1-2\sqrt{1-\widetilde{W}^{2}%
}\right)  $. The covariant derivative induced via pullback (it is necessary to
make the correspondence with the electroweak sector of the SM) can be
immediately identified~\cite{Fairlie:1979at,Weinberg:1967tq}
\begin{equation}
D_{\mu}\widetilde{\phi}=\partial_{\mu}\widetilde{\phi}-2i\left[  \sigma
\cdot\widetilde{W}_{\mu}-\frac{1}{\sqrt{3}}\mathbb{I}_{2\times2}\widetilde
{B}_{\mu}\left(  1-2\sqrt{1-\widetilde{W}^{2}}\right)  \right]  \widetilde
{\phi}.
\end{equation}
The electroweak mixing angle $\theta_{W}$ is determined via the relation
between the gauge couplings $g$ and $g^{\prime}$ in the above equation:
\begin{equation}
D_{\mu}\phi=\partial_{\mu}\widetilde{\phi}-2i\left(  g\widetilde{W}_{\mu}%
\cdot\sigma-g^{\prime}\mathbb{I}_{2\times2}\widetilde{B}_{\mu}\right)
\widetilde{\phi},
\end{equation}
then
\begin{equation}
\sin^{2}\theta_{W}=\frac{g\prime^{2}}{g^{2}+g\prime^{2}}\rightarrow
\frac{\left(  1-2\sqrt{1-\widetilde{W}^{2}}\right)  ^{2}}{3+\left(
1-2\sqrt{1-\widetilde{W}^{2}}\right)  ^{2}}.
\end{equation}
If $W^{2}\sim0$ then
\begin{equation}
\left.  \sin^{2}\theta_{W}\right\vert _{W^{2}\rightarrow0}=\frac{g\prime^{2}%
}{g^{2}+g\prime^{2}}\rightarrow\frac{\left(  1-2\right)  ^{2}}{3+\left(
1-2\right)  ^{2}}\rightarrow0.25.
\end{equation}

\subsection{Even supercurvature}

In the same way as before for the odd part of the supercurvature, we get the
even part in the form
\begin{equation}
d\Gamma_{even}+\left[  \Gamma_{even},\Gamma_{even}\right]  +\left\{
\Gamma_{odd},\Gamma_{odd}\right\}  \rightarrow
\end{equation}%
\[
\rightarrow\left(
\begin{array}
[c]{cc}%
d\widetilde{W}_{k}+\widetilde{W}_{i}\wedge\widetilde{W}_{j}\epsilon
_{ijk}\sigma_{k}+d\left(  \frac{1}{\sqrt{3}}\widetilde{B}\right)
-4\widetilde{\phi}\widetilde{\phi^{\dagger}} & \mathbf{0}_{2\times1}\\
\mathbf{0}_{1\times2} & d\left(  \frac{2}{\sqrt{3}}\widetilde{B}\right)
-4\widetilde{\phi^{\dagger}}\widetilde{\phi}%
\end{array}
\right) .
\]
Consequently, the full supercurvature takes the form
\begin{equation}
\mathcal{F}\equiv\left(
\begin{array}
[c]{cc}%
d\widetilde{W}_{k}+\widetilde{W}_{i}\wedge\widetilde{W}_{j}\epsilon
_{ijk}\sigma_{k}+d\left(  \frac{1}{\sqrt{3}}\widetilde{B}\right)
-4\widetilde{\phi}\widetilde{\phi^{\dagger}} & D\widetilde{\phi}\\
\left(  D\widetilde{\phi}\right)  ^{\dagger} & d\left(  \frac{2}{\sqrt{3}%
}\widetilde{B}\right)  -4\widetilde{\phi^{\dagger}}\widetilde{\phi}%
\end{array}
\right)  .
\end{equation}
Note that a tilde indicates here that the respective quantities are affected
by the induced supercurvature due the pullback from the algebra (vector space)
to the group representation.

\section{Superconnections and supergeometry}

The strategy to extend the symmetry without breaking the group theoretical
features of the model is realized as follows.

i) If we have two diffeomorphic (or gauge) non-equivalent $SU(2|1)$ valuated
superconnections, namely $\Gamma^{AB}$ and $\widetilde{\Gamma}^{AB}$. Their
difference transforms as a second rank three-supertensor under the action of
$SU(2|1)$:
\begin{align}
\kappa^{AB}  &  =G_{\text{ }C}^{A}G_{\text{ }D}^{B}\kappa^{CD},\\
\kappa^{AB}  &  \equiv\widetilde{\Gamma}^{AB}-\Gamma^{AB}.
\end{align}

ii) If we calculate now the curvature from $\widetilde{\Gamma}^{AB}$, we
obtain
\begin{equation}
\widetilde{R}^{AB}=R^{AB}+\mathcal{D}\kappa^{AB},
\end{equation}
where the $SU(2|1)$ supercovariant derivative is defined in the usual way (see
the previous Section)
\begin{equation}
\mathcal{D}\kappa^{AB}=d\kappa^{AB}+\Gamma_{\text{ }C}^{A}\wedge\kappa^{CB}
+ \Gamma_{\text{ }D}^{B}\wedge\kappa^{AD}.
\end{equation}

iii) Redefining the $SU(2|1)$ three vectors as $V_{2}^{A}\equiv\psi^{A}$ and
$V_{1}^{B}\equiv\varphi^{B}$ (in order to put all in the standard notation),
the 2-form $\kappa^{AB}$ can be constructed as
\begin{equation}
\kappa^{AB}\rightarrow\psi^{\left[  A\right.  }\varphi^{\left.  B\right\}
}dU,
\end{equation}
where $U$\ is a super-scalar function. Then we introduce all into the
$\widetilde{R}^{AB}$ and get
\begin{align}
\widetilde{R}^{AB}  &  =R^{AB}+\mathcal{D}\left(  \psi^{\left[  A\right.
}\varphi^{\left.  B\right\}  }dU\right) \nonumber\\
&  =R^{AB}+\left(  \psi^{\left[  A\right.  }\mathcal{D}\varphi^{\left.
B\right\}  }-\varphi^{\left[  A\right.  }\mathcal{D}\psi^{\left.  B\right\}
}\right)  \wedge dU.
\end{align}
Note that the supercurvature $\widetilde{R}^{AB}$ splits into even and odd
parts as indicated in the previous Section, being the capital letters the
multi-index $A,B,C$ etc. corresponding to the supercoordinates of the
$su(2|1)$ superspace.

iv) Let us define
\begin{equation}
\widetilde{\theta}^{A}=\widetilde{\mathcal{D}}\varphi^{A} \label{tag:29}%
\end{equation}
with the extended superconnection $\widetilde{\Gamma}^{AB}=\Gamma^{AB}%
+\kappa^{AB}$, then
\begin{align}
\widetilde{\theta}^{A}  &  = \underset{\theta^{A}}{\underbrace{\mathcal{D}%
\varphi^{A}}}+\kappa_{\text{ }B}^{A}\varphi^{B},\nonumber\label{tag:30}\\
\widetilde{\theta}^{A}  &  = \theta^{A}+\left[  \psi^{A}\left(  
\varphi^{B}\right)^{2} - \varphi^{A}\left( \psi\cdot\varphi\right)  \right]  
\wedge dU,
\end{align}
where $\left(  \varphi^{B}\right)  ^{2}=\left(  \varphi_{\text{ }B}\varphi
^{B}\right)  $ and $\left(  \psi\cdot\varphi\right)  =\psi_{B}\varphi^{B}$ etc.

In the same manner we also define
\begin{align}
\widetilde{\eta}^{A}  &  =\widetilde{\mathcal{D}}\psi^{A},\nonumber\\
\widetilde{\eta}^{A}  &  =\eta^{A}+\left[  \psi_{2}^{A}\left(  \psi
\cdot\varphi\right)  -\varphi^{A}\left(  \psi^{B}\right)  ^{2}\right]  \wedge
dU.
\end{align}

\section{Extended supercurvature}

From the original superconnection extended in the way described in the
previous Section, the extended supercurvature is computed in the
straightforward manner as
\begin{align}
&  \mathcal{F}_{ext}^{AB}=\mathcal{F}^{AB}+\mathcal{D}_{e}\left(
\psi^{\left[  A\right.  }\varphi^{\left.  B\right\}  }dU\right) \\
&  \qquad=\mathcal{F}^{AB}+\left(  \psi^{\left[  A\right.  }\mathcal{D}%
\varphi^{\left.  B\right\}  }-\varphi^{\left[  A\right.  }\mathcal{D}%
\psi^{\left.  B\right\}  }\right)  \wedge dU\nonumber\\
&  \qquad+\left(  \psi^{\left[  A\right.  }\kappa_{C}^{B}\varphi^{\left.
C\right\}  }-\varphi^{\left[  A\right.  }\kappa_{C}^{B}\psi^{\left.
C\right\}  }\right)  \wedge dU\\
&  \qquad=\left(
\begin{array}
[c]{cc}%
D\widetilde{W}_{k}+\frac{1}{\sqrt{3}}D\widetilde{B}-4\widetilde{\phi
}\widetilde{\phi^{\dagger}} & D\widetilde{\phi}-2\widetilde{\psi}%
\widetilde{\phi}\widetilde{\varphi}\\
\left(  D\widetilde{\phi}\right)  ^{\dagger}+2\widetilde{\varphi}%
\widetilde{\phi}^{\dagger}\widetilde{\psi} & \frac{2}{\sqrt{3}}D\widetilde
{B}-4\widetilde{\phi^{\dagger}}\widetilde{\phi}+\left(  \widetilde{\psi
}d\widetilde{\varphi}-\widetilde{\varphi}d\widetilde{\psi}\right)
+2\widetilde{\psi}^{2}\widetilde{\varphi}^{2}%
\end{array}
\right)  , \label{Fext}%
\end{align}
where $2\left(  \widetilde{\psi}^{2}\widetilde{\varphi}^{2}+\left(
\widetilde{\psi}\cdot\widetilde{\varphi}\right)^{2}\right)  =\widetilde
{\psi}^{2}\widetilde{\varphi}^{2}$ due to the $N=1$ Fierz 
identities\footnote{for $N=1$ we use
$\left(\psi\varphi\right) \left(\psi\varphi\right)
=-\frac{1}{2}\psi^{2}\varphi^{2}$} we get
$D\widetilde{W}_{k}=\left( d\widetilde{W}_{k}
+\widetilde{W}_{i}\wedge\widetilde{W}_{j}\epsilon_{i}jk\right)
\sigma_{k}$ and $D\widetilde{B}=d\widetilde{B}$. 
Note that, at the supercurvature level, a Dirac-type term, namely 
$\left(  \widetilde{\psi}d\widetilde{\varphi}
-\widetilde{\varphi}d\widetilde{\psi}\right)
+2\widetilde{\psi}^{2}\widetilde{\varphi}^{2}$ plus the couplings with
$\widetilde{\phi^{\dagger}}$ and $\widetilde{\phi}$ are geometrically
induced by the extension of the original superalgebra.

\section{Superlagrangian}

\subsection{Case 1: $\widetilde{\psi},\widetilde{\varphi}=const.$}

From Eq.~(\ref{Fext}) in the constant case for the fiducial vectors, namely for
$d\widetilde{\varphi}=d\widetilde{\psi}=0$, we have
\begin{align}
S  &  =\frac{1}{4}\left\langle \mathcal{F}_{ext},\mathcal{F}_{ext}%
\right\rangle \\
&  =\int d^{4}x\left[  -\frac{1}{4}\left(  \left(  F_{W}^{ext}\right)
_{\mu\nu}\left(  F_{W}^{ext}\right)  ^{\mu\nu}+\left(  F_{B}^{ext}\right)
_{\mu\nu}\left(  F_{B}^{ext}\right)  ^{\mu\nu}\right)  +\left(  D\widetilde
{\phi}\right)  ^{\dagger}D\widetilde{\phi}-V\left(  \widetilde{\phi}^{\dagger
},\widetilde{\phi},\widetilde{\psi},\widetilde{\varphi}\right)  \right]
,\nonumber
\end{align}%
\begin{equation}
F_{Wk}^{ext}=d\widetilde{W}_{k}+\widetilde{W}_{i}\wedge\widetilde{W}%
_{j}\epsilon_{ijk},
\end{equation}%
\begin{equation}
F_{B\mu\nu}^{ext}=\partial_{\mu}B_{\nu}-\partial_{\nu}B_{\mu},
\end{equation}%
\begin{equation}
V\left(  \widetilde{\phi^{\dagger},}\widetilde{\phi},\widetilde{\psi
},\widetilde{\varphi}\right)  =16\left[  \left(  \widetilde{\phi^{\dagger}%
}\widetilde{\phi}-\frac{v^{2}}{8}\right)  \left(  \widetilde{\phi^{\dagger}%
}\widetilde{\phi}+\frac{v^{2}}{8}\right)  \right]  , \label{V}%
\end{equation}
where we defined
\begin{equation}
v^{2}=2\left(  \widetilde{\psi}^{2}\widetilde{\varphi}^{2}+\left(
\widetilde{\psi}\cdot\widetilde{\varphi}\right)  ^{2}\right)  =\widetilde
{\psi}^{2}\widetilde{\varphi}^{2}. \label{v}%
\end{equation}
Gauge couplings $g$ and $g^{\prime}$ are not modified in the covariant
derivative. Consequently, assuming similar conditions as in the SM (see that
the even sector is just $SU(2)\otimes U(1)$), so the masses of the
$\widetilde{W}$ and $\widetilde{Z}=\frac{\sqrt{3}\widetilde{W}^{3}%
-\widetilde{B}}{2}$ gauge bosons become
\begin{equation}
M_{\widetilde{W}}=g\frac{v}{2},\qquad M_{\widetilde{Z}}=\sqrt{g^{2}%
+g^{\prime2}}\;\frac{v}{2}.
\end{equation}
For $g=1$ the expressions are simplified even more.

\subsection{Case 2: $\widetilde{\psi}=\widetilde{\psi}\left(  z\right)
,\widetilde{\varphi}=\widetilde{\varphi}\left(  z\right)  $}

In this particular case we propose
\begin{equation}
\left(  \widetilde{\psi}d\widetilde{\varphi}-\widetilde{\varphi}%
d\widetilde{\psi}\right)  +\widetilde{\psi}^{2}\widetilde{\varphi}^{2}\equiv
v^{2}. \label{econd}%
\end{equation}
Consequently, multiplying~(\ref{econd}) by 
$\left(\widetilde{\psi}\widetilde{\varphi}\right)^{-1}$ 
we obtain the differential equation for the
contraction $\left(\widetilde{\psi}\widetilde{\varphi}\right)$ and get the
following result
\begin{equation}
\left(  \widetilde{\psi}\widetilde{\varphi}\right)  =\underset{\varrho
}{\underbrace{\left(  \widetilde{\psi_{0}}\widetilde{\varphi_{0}}\right)  }%
}\frac{v}{\sqrt{2}}\tanh[(z-z_{0})\sqrt{2}v]. \label{kink}%
\end{equation}
Note that the equation is on the third components of the respective fiducial
vectors $\psi^{A}$\ and $\varphi^{B}$and $z$\ is a super coordinate of the
manifold $SU(2\mid1)$\ -valuated. Here $z\equiv z^{A}E_{A}$ is a general
coordinate of the supergeometry (induced via the pullback with 
$\left\{E_{A}\right\}$ the corresponding vector basis) from $su(2\mid1))$. 
The above result is new and is very important due to the possibility 
to change the character of the $su(2\mid1)$ valuated supercovariant 
derivative into a dynamical one.
\begin{figure}
[ptb]
\begin{center}
\includegraphics[
natheight=2.044400in,
natwidth=3.186000in,
height=2.0825in,
width=3.2309in
]%
{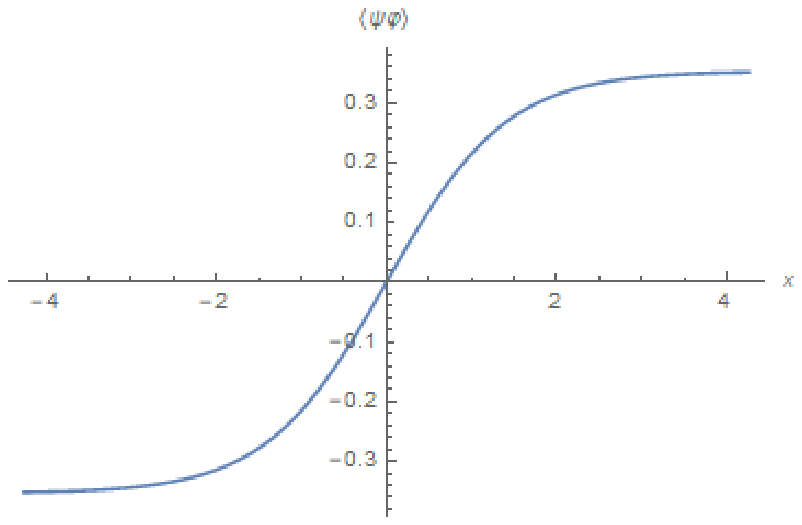}%
\caption{ $\left(  \widetilde{\psi}\widetilde{\varphi}\right)  $ for $v=0.5$.}%
\label{f1}%
\end{center}
\end{figure}
\begin{figure}
[ptb]
\begin{center}
\includegraphics[
natheight=2.061700in,
natwidth=3.186000in,
height=2.1006in,
width=3.2309in
]%
{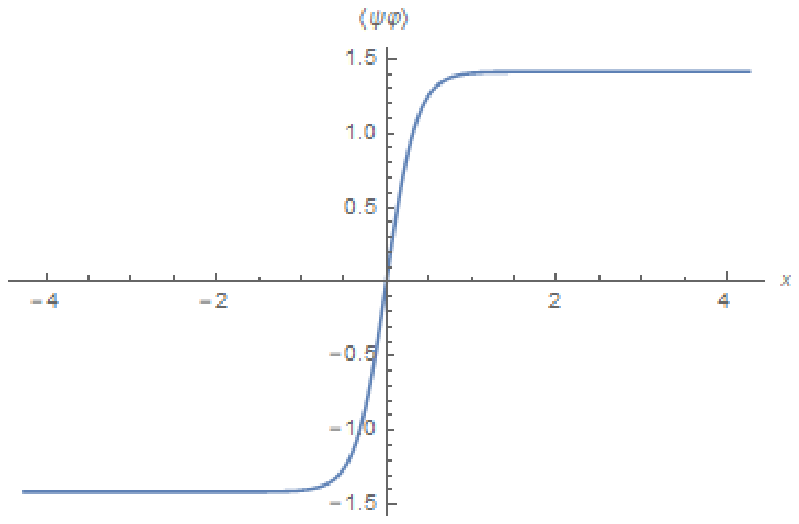}%
\caption{ $\left(  \widetilde{\psi}\widetilde{\varphi}\right)  $ for $v=2$.}%
\label{f2}%
\end{center}
\end{figure}
\begin{figure}
[ptb]
\begin{center}
\includegraphics[
natheight=2.044400in,
natwidth=3.186000in,
height=2.0825in,
width=3.2309in
]%
{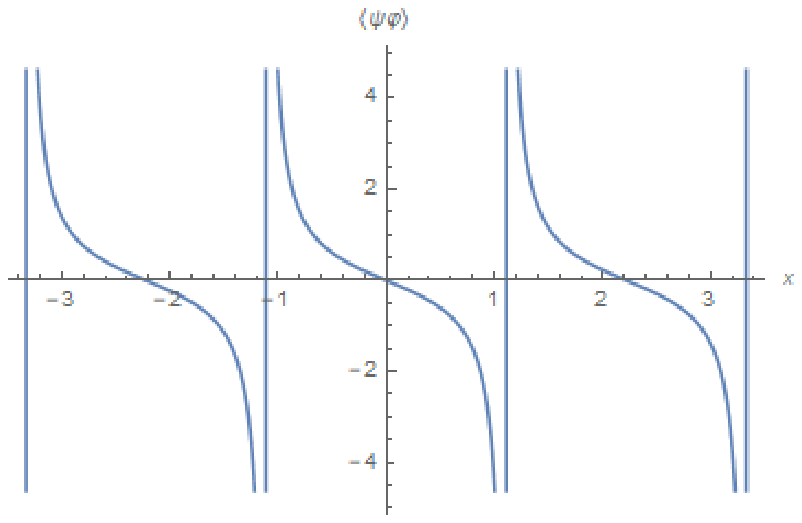}%
\caption{$\left(\widetilde{\psi}\widetilde{\varphi}\right)$ for $v=-i$.}%
\label{f3}%
\end{center}
\end{figure}
\begin{figure}
[ptb]
\begin{center}
\includegraphics[
natheight=2.044400in,
natwidth=3.186000in,
height=2.0825in,
width=3.2309in
]%
{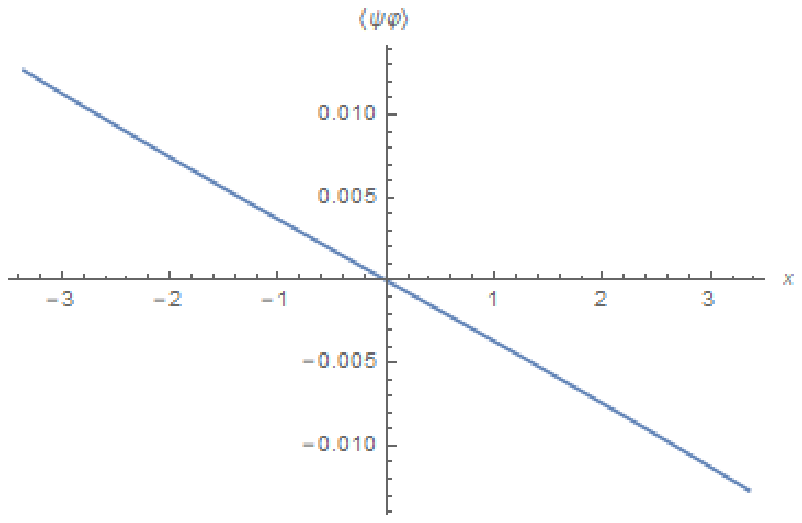}%
\caption{$\left(\widetilde{\psi}\widetilde{\varphi}\right)$ for $v=0.6i$.}%
\label{f4}%
\end{center}
\end{figure}
Then, the extended supercurvature becomes 
\begin{equation}
\left(
\begin{array}
[c]{cc}%
D\widetilde{W}+D\left(  \frac{1}{\sqrt{3}}\widetilde{B}\right)  -4\widetilde
{\phi}\widetilde{\phi^{\dagger}} & D\widetilde{\phi}-\sqrt{2}\varrho
v\tanh[(z-z_{0})\sqrt{2}v]\widetilde{\phi}\\
\left(  D\widetilde{\phi}\right)  ^{\dagger}+\sqrt{2}\varrho v\tanh
[(z-z_{0})\sqrt{2}v]\widetilde{\phi}^{\dagger} & D\left(  \frac{2}{\sqrt{3}%
}\widetilde{B}\right)  -4\widetilde{\phi^{\dagger}}\widetilde{\phi}+v^{2}%
\end{array}
\right). \label{dynC}%
\end{equation}%
\begin{figure}
[ptb]
\begin{center}
\includegraphics[
natheight=2.403300in,
natwidth=3.628800in,
height=3.3658in,
width=5.0609in
]%
{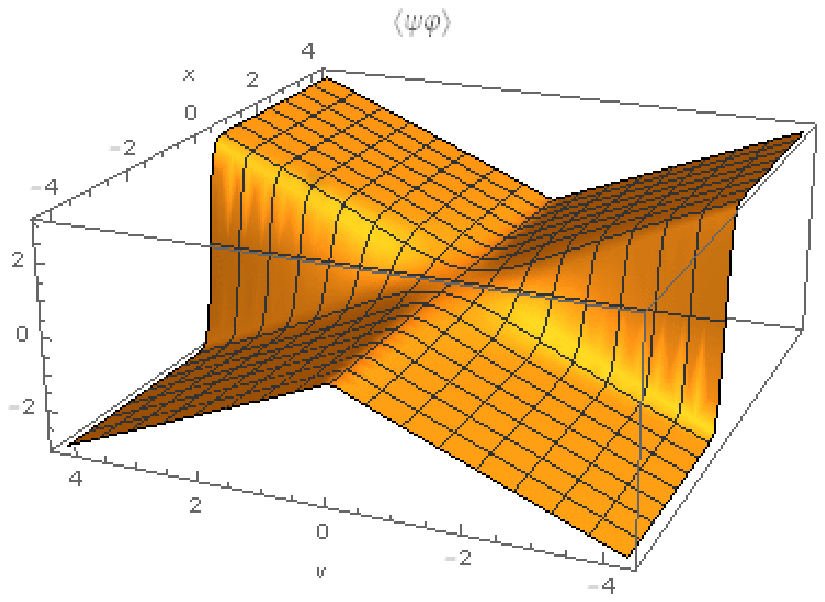}%
\caption{$\left(  \widetilde{\psi}\widetilde{\varphi}\right)  $ as a function
of spacetime coordinate and $v$.}%
\label{f5}%
\end{center}
\end{figure}
\begin{figure}
[ptb]
\begin{center}
\includegraphics[
natheight=2.254600in,
natwidth=5.469100in,
height=2.744in,
width=6.6098in
]%
{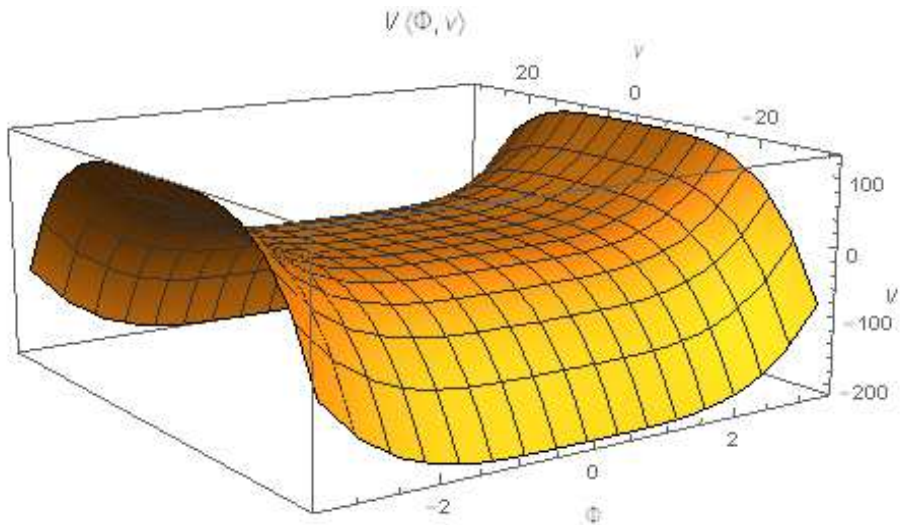}%
\caption{Superpotential as function of $\phi$ and $v$.}%
\label{f6}%
\end{center}
\end{figure}

As in the standard case, we can take $\phi$ adimensionalized as
appear in the exponential representation of the $SU(2\mid1)$ (from the group
to the physical scenario). To this end we construct the $\phi$ field as
\begin{equation}
\phi=\frac{1}{\phi_{0}}\left(
\begin{array}
[c]{c}%
\pi^{+}\\
\frac{v+h+i\pi^{0}}{2}%
\end{array}
\right)  \frac{1}{\sqrt{1+\frac{1}{\phi_{0}^{2}}\left(  \pi^{+2}+\frac{\left(
v+h\right)  ^{2}+\left(  \pi^{0}\right)  ^{2}}{4}\right)  }}%
\end{equation}
including the values $v$, $h$ for the Higgs field and $\pi$ and some bare
quantity $\phi_{0}$ to be determined. Extracting from the geometrically
induced extended superpotential $V\left(  \widetilde{\phi^{\dagger},}
\widetilde{\phi},\widetilde{\psi},\widetilde{\varphi}\right)$, 
see Eq.~(\ref{V}), the (adimensionalized) mass term for $h$
\begin{equation}
\sim\cdot\cdot\cdot+\frac{1}{\phi_{0}^{4}}\frac{16\left(  2vh\right)  ^{2}%
}{4\left[  1+\frac{1}{\phi_{0}^{2}}\left(  \pi^{+2}+\frac{\left(  v+h\right)
^{2}+\left(  \pi^{0}\right)  ^{2}}{4}\right)  \right]  ^{2}}+\cdot\cdot\cdot
\end{equation}
Then, at the tree level we obtain the \textit{adimensionalized} Higgs mass as%
\begin{equation}
\left.  M_{H}\right\vert _{treeAD}\sim\frac{4\sqrt{2}v}{\phi_{0}\left[
1+\frac{1}{\phi_{0}^{2}}\left(  \pi^{+2}+\frac{\left(  v+h\right)
^{2}+\left(  \pi^{0}\right)  ^{2}}{4}\right)  \right]  }.
\end{equation}
Consequently the \textit{physical} mass is given by%
\begin{equation}
\left.  M_{H}\right\vert _{tree}=\phi_{0}\left.  M_{H}\right\vert
_{treeAD}\sim\frac{4\sqrt{2}v}{\left[  1+\frac{1}{\phi_{0}^{2}}\left(
\pi^{+2}+\frac{\left(  v+h\right)  ^{2}+\left(  \pi^{0}\right)  ^{2}}%
{4}\right)  \right]  }\, .%
\end{equation}
The normalizing field $\phi_{0}$ can be determined from the
expression~(\ref{kink}) by the constant of integration that would indicate
that $\rho^{-1}\sim\phi_{0}.$

\section{Symmetry breaking mechanism}

In spite of great theoretical achievements on can't say that symmetry breaking
(spontaneous or not) in gauge field theories and nonlinear phenomenological
Lagrangians is completely well understood. A rather close relationship of
spontaneous symmetry breaking with nonlinearly realized symmetries was
stressed long time ago by many people~\cite{ColWess,VA,BO,SamWess,IK}. 
In the formalism of nonlinear realizations, we are
dealing with the invariance of a Lagrangian under a given group $G$ of field
transformations. Generally these invariances, however, are partially linear
and partially nonlinear. It happens that there is a subgroup $H$ of $G$,
called the stability group, which in well studied spontaneously broken
theories corresponds to the symmetry group of the vacuum state and acts on all
fields and the covariant derivatives linearly. The remaining transformations
of the larger group of invariance, namely, those belonging to the coset $G/H$,
change the vacuum state and produce nonlinear transformations on the physical
fields and covariant derivatives.

In this context, following our previous work~\cite{Arbuzov:2017frk,dc}, the
natural interpretation in the language of coherent states is extremely clear
and concise, namely the group structure is chosen in the form
\begin{equation}
G=G/H\otimes H.\nonumber
\end{equation}
The subgroup $H$ gives the transformations defining the vacuum states and
consequently the set of fiducial vectors spanning the vacuum sub-space.

The coset $G/H$ gives the remaining transformations changing the vacuum states
and therefore producing nonlinear (Bogoliubov) transformations of physical
fields and induced covariant derivatives. As we can see from the previous
Sections, the breaking of the supersymmetries in this case is driven by the
supercoherent states (constructed by the action of the supergroup $SU(2\mid1)$
on supervectors as described in Appendixes~I and III) in the same way as in
Ref.~\cite{Arbuzov:2017frk}. However, in this case, the original
representation is still preserved because everything is an internal space of
symmetry: it only extends in contrast to our previous work where the
representation also included the external space-time.

\section{Concluding remarks and outlook}

In this paper we developed a possible description of the electroweak sector of
the SM using the methods described in our paper~\cite{Arbuzov:2017frk}. As we
see, using naturally a coherent superstate based on the simplest supergroup
$SU(2\mid1)$ which is the group of dynamic symmetry of the supersphere. This
is naturally isomorphic to the group $OSp(2\mid2)$ from the point of view of
algebra, keeping invariant in the natural factorization, the even part
$SU(2)\otimes U(1)$. Our interpretation of the odd sector is that physically
it might be a hidden counterpart of the Standard Model.

In one of the cases discussed here that is particularly interesting, the
diagonal part corresponding to the even sector that defines a geometrically
induced differential equation for a component of a super-field. This
differential equation can be considered non-homogeneous, specifically equal to
the constant of the mean value of the condensate $v^{2}$ or superscalar
product between such coherent superfields that extend the original
superalgebra. The resolution of the differential equation results in a
solution \ref{dynC} as indicated in figures \ref{f1}---\ref{f5}. They
describes the compact solutions (\ref{f1},\ref{f2}) and noncompact ones
(\ref{f3},\ref{f4}) respectively. Fig.~\ref{f5} represents the 3-dimensional
picture and Fig.~\ref{f6} the superpotential. This fact, namely getting the
supersymmetric solitonic solutions~\cite{Mshif} shows the superintegrability
of the super-extended model.

It is good to note that the superpotential is an exact difference of squares
in contrast to the standard case where the quadratic part appears. With
respect to the statistics and other issues corresponding to the structure of
the group, here the field corresponding to the Higgs appears as odd in the
representation but as $SU(2)$ scalar (doublet) from the Lagrangian point of
view. In this same line we see that the field associated with the
Higgs must be a scalar of $SU(2)$ because it is extended by means of the
"fermionic condensate" given by the fiducial vectors (vacuum, $\psi\varphi$)
of the supercoherent states (see Figs.~(\ref{V},\ref{v})). Consequently, there
is no problem with respect to the bosonic character of $\phi$. However, the
very geometrical structure of the simplest supergroup must define the
correctness of the sign of the dynamical terms of the additional superfields.

This fact (under investigation now) is in part connected with the observation
made in Ref.~\cite{Fairlie:1979at}: changing the value of $\lambda_{8}$ to
have trace equal to $-2$, the structure of the representation goes from
$SU(3)$ to $SU(2\mid1)$, i.e. from a Lie structure to a
graded Lie one. In our case, the final structure is of a dynamic character.

Regarding the relationship with the approach of nonlinear realizations, the
link is direct considering that the field $\widetilde{\phi}$\ which plays the
role of Higgs could be clearly eliminated at the expense of the fields of the
hidden super-sector $\widetilde{\psi},\widetilde{\varphi}$\ and the constant
$v$(playing the final role of expectation value). This observation is
consistent considering that the antidiagonal part of the supercurvature
(\ref{dynC}) it is precisely the Maurer-Cartan superform associated with the
breaking of some (super) symmetry in a standard way: canceling precisely that
superform (e.g. $D\widetilde{\phi}=d\widetilde{\phi})$%
\begin{equation}
\omega_{\widetilde{\phi}}=0\rightarrow d\widetilde{\phi}=\sqrt{2}\varrho
v\tanh[(z-z_{0})\sqrt{2}v]\widetilde{\phi}\rightarrow d\left(  \ln
\widetilde{\phi}\right)  =\varrho\tanh X\text{ \ }dX,\label{nn1}%
\end{equation}
with $X\equiv(z-z_{0})\sqrt{2}v$, the result is easily obtained as%
\begin{equation}
\widetilde{\phi}=\widetilde{\phi}_{0}(\cosh[(z-z_{0})\sqrt{2}v])^{\rho}\text{
\ ,\ \ \ \ \ }\widetilde{\phi}_{0}=const\text{\ \ }\label{nn2}%
\end{equation}
being $z\equiv z^{A}E_{A}$\ the supercoordinate $SU(2\mid1)$\ valuated (with
the superbasis carrying the superalgebra symmetry). If we define again
$\tanh\left(  X/2\right)  =\Lambda$\ then (\ref{nn2}) can be written
as%
\begin{equation}
\widetilde{\phi}=\widetilde{\phi}_{0}\left(  \frac{1+\Lambda^{2}}%
{1-\Lambda^{2}}\right)  ^{\rho}\text{ \ ,\ \ \ \ \ }\widetilde{\phi}%
_{0}=const\text{\ \ }\label{nn3}%
\end{equation}
The above results are very important:

1) the "Higgs"\ doublet $\widetilde{\phi}$\ as we see is soliton type
superintegrable for determined values of $\rho<0$\ (our "fermion condensate")
and for the other case it shape turns to be divergent as we see in figure in
this Section.%

\begin{figure}
[ptb]
\begin{center}
\includegraphics[
natheight=1.991700in,
natwidth=3.088200in,
height=2.0297in,
width=3.1332in
]%
{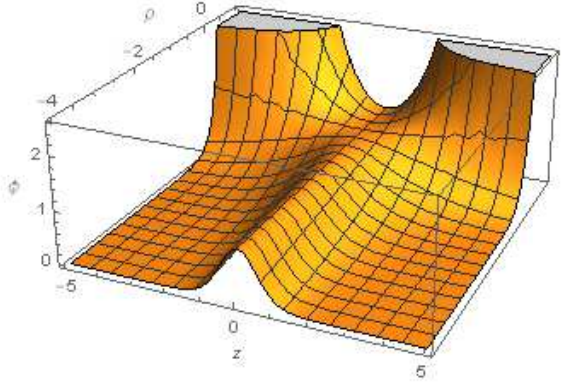}%
\end{center}
\caption{The shape of the Higgs doublet $phi$ as a function of the $su(2|1)$ 
valuated coordinates and fermion condensate $\rho$.}
\end{figure}

2) When $\rho$\ is zero, the cut branch appears, passing the field
$\widetilde{\phi}$from a compact configuration (soliton type) to a divergent one.

3) We can see from (\ref{nn3}) that $\widetilde{\phi}$\ depends on the fields
$\widetilde{\psi},\widetilde{\varphi}$\ of the hidden supersector of the model.

4)From the dynamic point of view the difference with the solutions in the
similar cases obtained in other approaches (e.g. nonlinear realizations) is
the exponential dependence with respect to $\rho$\ that gives the soliton
shape (location / confinement) of the Higgs doublet.

In the next step, explicit form of the $\psi,\varphi$ supercoherent states
will be performed and the full supergroup $SU(2,2\mid1)$ will be considered in
the same way as a basis of a model containing the extension of GR\ and SM in a
unified way.

\section*{Acknowledgments}
D.J.C.-L. is grateful to the Bogoliubov Laboratory of Theoretical Physics JINR
for hospitality and CONICET-Argentina for financial support.

\section{Appendix I. Coset coherent states}

Let us remind the definition of coset coherent states
\begin{equation}
H_{0}=\left\{  g\in G\mid\mathcal{U}\left(  g\right)  V_{0}=V_{0}\right\}
\subset G. \label{tag:1}%
\end{equation}
Consequently the orbit is isomorphic to the coset, i.e.
\begin{equation}
\mathcal{O}\left(  V_{0}\right)  \simeq G/H_{0}. \label{tag:2}%
\end{equation}
Analogously, if we remit to the operators
\begin{equation}
\left\vert V_{0}\right\rangle \left\langle V_{0}\right\vert \equiv\rho_{0}.
\label{tag:3}%
\end{equation}
Then the orbit
\begin{equation}
\mathcal{O}\left(  V_{0}\right)  \simeq G/H \label{tag:4}%
\end{equation}
with
\begin{align}
H  &  =\left\{  g\in G\mid\mathcal{U}\left(  g\right)  V_{0}=\theta
V_{0}\right\} \nonumber\label{tag:5}\\
&  =\left\{  g\in G\mid\mathcal{U}\left(  g\right)  \rho_{0}\mathcal{U}%
^{\dagger}\left(  g\right)  =\rho_{0}\right\}  \subset G.
\end{align}
The orbits are identified with cosets spaces of $G$ with respect to the
corresponding stability subgroups $H_{0}$ and $H$ being the vectors $V_{0}$ in
the second case defined within a phase. From the quantum viewpoint $\left\vert
V_{0}\right\rangle \in\mathcal{H}$ (the Hilbert space) and $\rho_{0}%
\in\mathcal{F}$ (the Fock space) are $V_{0}$ normalized fiducial vectors
(embedded unit sphere in $\mathcal{H}$).

\section{Appendix II. Superconnection and algebra generating the group}

i) The superconnection as defined in the case of \cite{Fairlie:1979at,Neeman:2005dbs} 
involves the vector space via gauging over the
superalgebra where the even part is, phenomenologically speaking, as in the
electroweak sector
\[
\mathcal{J}=i\left[
\begin{array}
[c]{cc}%
\mathcal{W}-\frac{1}{\sqrt{3}}B\cdot\mathbb{I}_{2\times2} & \mathbf{0}%
_{2\times1}\\
\mathbf{0}_{1\times2} & -\frac{2}{\sqrt{3}}B
\end{array}
\right]  +i\left[
\begin{array}
[c]{cc}%
\mathbf{0}_{2\times2} & \sqrt{2}\phi\\
\sqrt{2}\phi^{\dagger} & 0
\end{array}
\right]  ,\qquad\mathcal{W\equiv}W\cdot\sigma.
\]

ii) In our case, we are working over the group structure (pullback) where the
physics lives
\[
\mathcal{U}=\frac{e^{-i\theta_{8}/\sqrt{3}}}{\sqrt{1+Z^{\dagger}Z}}\left\{
\left[
\begin{array}
[c]{cc}%
w & \mathbf{0}_{2\times1}\\
\mathbf{0}_{1\times2} & e^{-i\theta_{8}/\sqrt{3}}%
\end{array}
\right]  +\left[
\begin{array}
[c]{cc}%
\mathbf{0}_{2\times2} & \phi e^{-i\theta_{8}/\sqrt{3}}\\
\phi^{\dagger}w & \mathbf{0}%
\end{array}
\right]  \right\}  .
\]
The difference with respect to the previous case is evident.

3) From the algebra generating the group (e.g. $\mathcal{J\rightarrow U)}$ via
the pullback we obtain the nonlinear structure of fields as follows
\begin{gather}
\mathcal{W\rightarrow}\widetilde{\mathcal{W}}\mathcal{\equiv}\widetilde
{W}\cdot\sigma\rightarrow\frac{e^{-i\theta_{8}/\sqrt{3}}} {\sqrt
{1+\phi^{\dagger}\phi}}\frac{i}{\sqrt{1+W^{2}}}\left(
\begin{array}
[c]{cc}%
W_{3} & W_{-}\\
W_{+} & -W_{3}%
\end{array}
\right) \\
-\frac{1}{\sqrt{3}}B\cdot\mathbb{I}_{2\times2}\rightarrow-\frac{1}{\sqrt{3}%
}\widetilde{B}\cdot\mathbb{I}_{2\times2}\equiv-\frac{e^{-i\theta_{8}/\sqrt{3}%
}}{\sqrt{1+\phi^{\dagger}\phi}}\frac{i}{\sqrt{1+W^{2}}}\mathbb{I}_{2\times2}
\\
\text{ \ \ \ \ \ (note that }\quad \frac{1}{\sqrt{1+\phi^{\dagger}\phi}}%
=\cos\left\vert v\right\vert ),
\end{gather}%
\begin{align}
W_{3}  &  \rightarrow\widetilde{W}_{3}\equiv W_{3}\frac{e^{-i\theta_{8}%
/\sqrt{3}}}{\sqrt{1+\phi^{\dagger}\phi}}\frac{i}{\sqrt{1+W^{2}}} \, ,\\
W_{\pm}  &  \rightarrow\widetilde{W}_{\pm}\equiv W_{\pm}\frac{e^{-i\theta
_{8}/\sqrt{3}}}{\sqrt{1+\phi^{\dagger}\phi}}\frac{i}{\sqrt{1+W^{2}}}\, ,%
\end{align}%
\begin{align}
\sqrt{2}\phi &  \rightarrow\sqrt{2}\widetilde{\phi}\equiv e^{-2i\theta
_{8}/\sqrt{3}}\frac{1}{\sqrt{1+\phi^{\dagger}\phi}}\underset{Z}{\underbrace
{\left(
\begin{array}
[c]{c}%
\theta_{4}+i\theta_{5}\\
\theta_{6}+i\theta_{7}%
\end{array}
\right)  }},\\
\sqrt{2}\overline{\phi}  &  \rightarrow\sqrt{2}\widetilde{\overline{\phi}%
}\equiv e^{-i\theta_{8}/\sqrt{3}}\frac{1}{\sqrt{1+\phi^{\dagger}\phi}%
}\underset{Z^{\dagger}}{\underbrace{\left(  \theta_{4}-i\theta_{5},\theta
_{6}-i\theta_{7}\right)  }}w\\
&  =\frac{ie^{-i\theta_{8}/\sqrt{3}}}{\sqrt{1+W^{2}}\sqrt{1+\phi^{\dagger}%
\phi}}\left(
\begin{array}
[c]{cc}%
W_{3}-i & W_{-}\\
W_{+} & -W_{3}-i
\end{array}
\right). \nonumber
\end{align}
Consequently, the explicit form of the even supercurvature is
\begin{align}
&  d\Gamma=d\left(  \frac{e^{-i\theta_{8}/\sqrt{3}}}{\sqrt
{1+\phi^{\dagger}\phi}}\right)  \wedge\left(
\begin{array}
[c]{cc}%
w & \mathbf{0}_{2\times1}\\
\mathbf{0}_{1\times2} & e^{-i\theta_{8}/\sqrt{3}}%
\end{array}
\right)  
+ \frac{e^{-i\theta_{8}/\sqrt{3}}}{\sqrt{1+\phi^{\dagger}\phi}}\left(
\begin{array}
[c]{cc}%
dw & \mathbf{0}_{2\times1}\\
\mathbf{0}_{1\times2} & d\left(  e^{-i\theta_{8}/\sqrt{3}}\right)
\end{array}
\right)  \\
&  +\frac{-1}{1+W^{2}}\left(  \frac{e^{-i\theta_{8}/\sqrt{3}}}{\sqrt
{1+\phi^{\dagger}\phi}}\right)  ^{2}\left(
\begin{array}
[c]{cc}%
W_{i}\wedge W_{j}\epsilon_{ijk}\sigma_{k} & \mathbf{0}_{2\times1}\\
\mathbf{0}_{1\times2} & 0
\end{array}
\right)  -4i\frac{e^{-i\theta_{8}/\sqrt{3}}}{\sqrt{1+\phi^{\dagger}\phi}%
}\left(
\begin{array}
[c]{cc}%
\phi e^{-i\theta_{8}/\sqrt{3}}\phi^{\dagger} & \mathbf{0}_{2\times1}\\
\mathbf{0}_{1\times2} & \phi^{\dagger}w\phi
\end{array}
\right)  \Gamma_{even}. \nonumber
\end{align}
The explicit odd part reads
\begin{gather}
d\Gamma_{odd}+\Gamma_{even}\wedge\Gamma_{odd} \equiv 
\nonumber \\
\equiv-i\Gamma_{even}\left\{  \left(
\begin{array}
[c]{cc}%
\mathbf{0}_{2\times2} & d\phi-2\left[  \widetilde{W}\cdot\sigma-\frac
{g^{\prime}}{\sqrt{3}}\widetilde{B}\cdot\mathbb{I}_{2\times2}\right]  \phi\\
d\phi^{\dagger}+2\phi^{\dagger}\left[  \widetilde{W}\cdot\sigma-\frac{1}%
{\sqrt{3}}\widetilde{B}\cdot\mathbb{I}_{2\times2}\left(  1-2\sqrt{1-W^{2}%
}\right)  \right]   & 0
\end{array}
\right)  \right. \nonumber  \\
\left.  +i\underset{}{\underbrace{\left(
\begin{array}
[c]{cc}%
\mathbf{0}_{2\times2} & -\phi\wedge d\ln\left(  \frac{2}{\sqrt{3}}%
\widetilde{B}\cdot\mathbb{I}_{2\times2}\right)   \\
\phi^{\dagger}\wedge d\ln\left(  \widetilde{W}\cdot\sigma-\frac{1}{\sqrt{3}%
}\widetilde{B}\cdot\mathbb{I}_{2\times2}\right)   & 0
\end{array}
\right)  }}\right\} .
\end{gather}
We have $g'=\left(  1-2\sqrt{1-W^{2}}\right)$, therefore this expression 
takes the form give in Section~IV.

\begin{remark}
Note that in the above redefined "tilde" quantities, the geometry of the group
manifold is included.
\end{remark}

\begin{remark}
Note that in the case of the $B$ field, it is induced by the phase $\theta
_{8}$, by $\left\vert \theta\right\vert $ (e.g. $W_{3},W_{\pm})$ and by the
supersymmetric sector of the model.
\end{remark}

\section{Appendix III. The extended superconnection: $\kappa^{AB}$}

First at all we denote here the supergeometric product for two supervectors as
usual: $\left(  V_{1}V_{2}\right)  \equiv V_{1}\cdot V_{2}+V_{1}\wedge V_{2}$,
in particular for the spinor parameters $\left(  \phi_{1}\phi_{2}\right)
=\phi_{1\alpha}\phi_{2}^{\alpha}+\phi_{1\alpha}\wedge\phi_{2\beta}$. Now we
construct $\kappa^{AB}$ as follows:

i) we select two fiducial spinors $\psi^{0},\varphi^{0}\in$ $SU(2|1)$ ground
state, that mean that are annihilated by the lower (upper) group operators as
defined in appendix

ii) $\psi^{0},\varphi^{0}$ under the action of the supergroup in the Borel
symmetrical representation take the following form%
\[
\psi=\pm\left(
\begin{array}
[c]{c}%
\Omega^{-1/4}\Phi_{\beta}^{\alpha}\psi_{a}^{0}\\
\Omega^{-1/4}\rho_{\beta}^{\alpha}\psi_{\alpha}^{0}\\
\pm\psi_{\beta}^{0}%
\end{array}
\right)  ,\varphi=\pm\left(
\begin{array}
[c]{c}%
\Omega^{\prime-1/4}\Phi_{\beta}^{\alpha\prime}\varphi_{a}^{0}\\
\Omega^{\prime-1/4}\rho_{\beta}^{\alpha\prime}\varphi_{a}^{0}\\
\pm\varphi_{\beta}^{0}%
\end{array}
\right)
\]

\bigskip where $\Omega=\left(  1-\overline{\rho}\rho+\omega^{\ast}%
\omega-\overline{\Phi}\Phi\right)  $ and primed parameters indicate that the
fiducial vectors are, in principle, under the action of different elements of
$SU(2|1)$:%
\[
SU(2|1)\ni\psi\wedge\varphi=\left(  \Omega\Omega^{\prime}\right)
^{-1/4}\left(
\begin{array}
[c]{ccc}%
\Phi^{\alpha\beta}\psi_{a}^{0}\wedge\Phi_{\beta}^{\gamma\prime}\varphi
_{\gamma}^{0} & \Phi_{\delta}^{\alpha}\psi_{a}^{0}\wedge\rho_{\beta}%
^{\gamma\prime}\varphi_{\gamma}^{0} & \pm\Omega^{\prime1/4}\Phi_{\delta
}^{\alpha}\psi_{a}^{0}\wedge\varphi_{\beta}^{0}\\
\rho_{\delta}^{\alpha}\psi_{\alpha}^{0}\wedge\Phi_{\beta}^{\gamma\prime
}\varphi_{\gamma}^{0} & \rho_{\delta}^{\alpha}\psi_{\alpha}^{0}\wedge
\rho_{\beta}^{\gamma\prime}\varphi_{\gamma}^{0} & \pm\Omega^{\prime1/4}%
\rho_{\delta}^{\alpha}\psi_{\alpha}^{0}\wedge\varphi_{\beta}^{0}\\
\pm\Omega^{1/4}\psi_{\delta}^{0}\wedge\Phi_{\beta}^{\gamma\prime}%
\varphi_{\gamma}^{0} & \pm\Omega^{1/4}\psi_{\delta}^{0}\wedge\rho_{\beta
}^{\gamma\prime}\varphi_{\gamma}^{0} & \psi_{\delta}^{0}\wedge\varphi_{\beta
}^{0}%
\end{array}
\right) .
\]
If the elements of the group are the same then%
\[
SU(2|1)\ni\psi\wedge\varphi=\Omega^{-1/2}\left(
\begin{array}
[c]{ccc}%
\Phi^{\alpha\beta}\psi_{a}^{0}\wedge\Phi_{\beta}^{\gamma}\varphi_{\gamma}^{0}
& \Phi_{\delta}^{\alpha}\psi_{a}^{0}\wedge\rho_{\beta}^{\gamma}\varphi
_{\gamma}^{0} & \pm\Omega^{\prime1/4}\Phi_{\delta}^{\alpha}\psi_{a}^{0}%
\wedge\varphi_{\beta}^{0}\\
\rho_{\delta}^{\alpha}\psi_{\alpha}^{0}\wedge\Phi_{\beta}^{\gamma}%
\varphi_{\gamma}^{0} & \rho_{\delta}^{\alpha}\psi_{\alpha}^{0}\wedge
\rho_{\beta}^{\gamma}\varphi_{\gamma}^{0} & \pm\Omega^{\prime1/4}\rho_{\delta
}^{\alpha}\psi_{\alpha}^{0}\wedge\varphi_{\beta}^{0}\\
\pm\Omega^{1/4}\psi_{\delta}^{0}\wedge\Phi_{\beta}^{\gamma}\varphi_{\gamma
}^{0} & \pm\Omega^{1/4}\psi_{\delta}^{0}\wedge\rho_{\beta}^{\gamma}%
\varphi_{\gamma}^{0} & \psi_{\delta}^{0}\varphi_{\beta}^{0}-\psi_{\beta}%
^{0}\varphi_{\delta}^{0}%
\end{array}
\right)
\]
and when the symmetry is broken changing the vacuum (inverse Bogoliubov
transformation) we obtain as expected%
\[
SU(2|1)\ni\kappa^{AB}\varpropto\Omega^{-1/2}\left(
\begin{array}
[c]{ccc}%
0 & 0 & 0\\
0 & 0 & 0\\
0 & 0 & \psi_{\delta}^{0}\varphi_{\beta}^{0}-\psi_{\beta}^{0}\varphi_{\delta
}^{0}%
\end{array}
\right).
\]

\section{Appendix IV. Determination of the ground state vectors $V_{0}$: the meaning of the physical vacuum}

i) In the pure boson $V_{0}$ case
\[
U_{+}\left(
\begin{array}
[c]{c}%
A\\
B\\
C
\end{array}
\right)  =0\rightarrow\left(
\begin{array}
[c]{ccc}%
0 & \omega_{u} & \varphi_{u}\\
0 & 0 & \rho_{u}\\
0 & 0 & 0
\end{array}
\right)  \left(
\begin{array}
[c]{c}%
A\\
B\\
C
\end{array}
\right)  =\left(
\begin{array}
[c]{c}%
0\\
0\\
0
\end{array}
\right)  \rightarrow V_{0}=\left(
\begin{array}
[c]{c}%
A\\
0\\
0
\end{array}
\right)
\]
and the coherent vector reads
\[
V=gV_{0}\rightarrow V_{0}=\frac{\left(  1-\overline{\rho}\rho\right)  }%
{\sqrt{\left(  1-\overline{\rho}\rho+\omega^{\ast}\omega-\overline{\Phi}%
\Phi\right)  }}\left(
\begin{array}
[c]{c}%
A\\
\mp\frac{-\omega^{\ast}+\overline{\Phi}\rho}{\left(  1-\overline{\rho}%
\rho\right)  }A\\
\frac{\overline{\Phi}\left(  1-\overline{\rho}\rho+\omega^{\ast}%
\omega-\overline{\Phi}\Phi\right)  ^{1/4}}{\left(  1-\overline{\rho}%
\rho\right)  }A
\end{array}
\right).
\]

ii) In the pure fermion $V_{0}$ case
\[
U_{-}\left(
\begin{array}
[c]{c}%
A\\
B\\
C
\end{array}
\right)  =0\rightarrow\left(
\begin{array}
[c]{ccc}%
0 & 0 & 0\\
-\omega_{l}^{\prime\ast} & 0 & 0\\
-\overline{\varphi_{l}}^{\prime} & -\overline{\rho}_{l}^{\prime} & 0
\end{array}
\right)  \left(
\begin{array}
[c]{c}%
A\\
B^{\ast}\\
\overline{C}%
\end{array}
\right)  =\left(
\begin{array}
[c]{c}%
0\\
0\\
0
\end{array}
\right)  \rightarrow V_{0}=\left(
\begin{array}
[c]{c}%
0\\
0\\
\overline{C}%
\end{array}
\right)
\]
and the coherent vector reads
\[
V=gV_{0}\rightarrow V_{0}=\left(
\begin{array}
[c]{c}%
\Phi\left(  1-\overline{\rho}\rho+\omega^{\ast}\omega-\overline{\Phi}%
\Phi\right)  ^{-1/4}\overline{C}\\
\rho\left(  1-\overline{\rho}\rho+\omega^{\ast}\omega-\overline{\Phi}%
\Phi\right)  ^{1/4}\overline{C}\\
\overline{C}%
\end{array}
\right).
\]

iii) In the boson-fermion (symmetric) $V_{0}$ case

\[
U_{s}\left(
\begin{array}
[c]{c}%
A\\
B\\
C
\end{array}
\right)  =0\rightarrow\left(
\begin{array}
[c]{ccc}%
0 & 0 & 0\\
-\omega_{l}^{\prime\ast} & 0 & \rho_{u}\\
-\overline{\varphi_{l}}^{\prime} & 0 & 0
\end{array}
\right)  \left(
\begin{array}
[c]{c}%
A\\
B^{\ast}\\
\overline{C}%
\end{array}
\right)  =\left(
\begin{array}
[c]{c}%
0\\
0\\
0
\end{array}
\right)  \rightarrow V_{0}=\left(
\begin{array}
[c]{c}%
0\\
B^{\ast}\\
0
\end{array}
\right)
\]
with the coherent vector
\[
V=gV_{0}\rightarrow V_{0}=\frac{\left(  1-\overline{\rho}\rho\right)  }%
{\sqrt{\left(  1-\overline{\rho}\rho+\omega^{\ast}\omega-\overline{\Phi}%
\Phi\right)  }}\left(
\begin{array}
[c]{c}%
\frac{\omega+\overline{\rho}\Phi}{\left(  1-\overline{\rho}\rho\right)  }\\
1\\
\frac{-\overline{\rho}}{\left(  1-\overline{\rho}\rho\right)  }%
\end{array}
\right)  B^{\ast}.
\]

The specific choice depends on the structure of the Fock space.

\end{document}